\documentclass[12pt]{revtex4}
\newcommand{\be}{\begin{equation}}
\newcommand{\ba}{\begin{eqnarray}}
\newcommand{\ee}{\end{equation}}
\newcommand{\ea}{\end{eqnarray}}
\newcommand{\cosech} { {\rm cosech}}

\begin{document}

\title{The Scattering amplitude for Newly found exactly solvable Potential}

\author{ Rajesh Kumar Yadav $^{a}$\footnote{e-mail address: rajeshastrophysics@gmail.com  }, 
Avinash Khare$^{b}$\footnote {e-mail address: khare@iiserpune.ac.in}
 and  {Bhabani Prasad Mandal$^{a}$}\footnote {e-mail address: bhabani.mandal@gmail.com}}
 \affiliation{$~^a$ Department of Physics,Banaras Hindu University,Varanasi-221005, INDIA.\\ 
$~^b$ Raja Ramanna Fellow, Indian Institute of Science Education (IISER) 
Pune 411021, INDIA.}

\vspace{1.5 in}

\begin{abstract}
 The scattering amplitude for the recently discovered exactly solvable 
shape invariant potential, which is isospectral to the generalized 
P\"oschl-Taylor potential, is calculated explicitly by considering the 
asymptotic behavior of the $X_{1}$ Jacobi exceptional polynomials associated 
with this system. 
\end{abstract}

\maketitle

\newpage


In recent years the ideas of Supersymmetric quantum mechanics (SQM) and
shape invariant potentials (SIP) have greatly enriched our understanding
of the exactly solvable potentials \cite{cks}. The search for the exactly
solvable potentials has been boosted greatly due to the recent discovery of 
exceptional orthogonal polynomials (EOP) (also known as $X_n$ Laguerre and 
$X_n$ Jacobi polynomials) \cite{eop2,eop3}. Unlike the usual orthogonal 
polynomials, these EOPs start with degree $n\geq 1$
 and still form a complete orthonormal set with respect to a positive definite 
innerproduct defined over a compact interval. This remarkable 
work lead Quesne \cite{que} to the discovery of two new SIP whose solution
is in terms of $X_1$ Laguerre and $X_1$ Jacobi polynomials. Subsequently, 
a third SIP was discovered whose solution is also in terms of $X_1$ Jacobi
polynomials \cite{bqr}. Subsequently, Odake and Sasaki constructed infinite
sets of new SIP corresponding to all these three cases whose eigenfunctions
are in terms of $X_n$ Laguerre and $X_n$ Jacobi polynomials \cite{os}. 

It is worth pointing out that unlike the usual SIP, the newly discovered SIP
are explicitly $\hbar$ dependent. Further, all of them are isospectral to the
well known SIP. Besides, while two out of the three newly
discovered SIP have pure bound state spectrum, the third SIP which is 
isospectral to the generalized P\"oschl-Teller (GPT) potential, has both
discrete and continuous spectrum. To the best of our knowledge, while the
bound state energy eigenvalues and eigenfunctions have already been obtained
in all the three cases, the scattering amplitude has still not been obtained
in the case of SIP which are isospectral to GPT. The purpose of this note
is to fill up this gap partially. In particular, 
in the present work we obtain the scattering amplitude for the newly 
discovered SIP whose solution is in terms of $X_1$ Jacobi polynomial. 

The superpotential corresponding to GPT (which is on the half line 
$ 0 \le r \le \infty$) is given by
\be\label{1}
W_{GPT} = A \coth r - B \cosech r ~~ B > A+1 > 1\,. 
\ee
The potential $V_{GPT}(r) = W^2_{GPT}(r) - W'_{GPT}(r)$ which follows from 
the above super potential is given by ($\hbar = 2m =1$)   
\be\label{2}
V_{GPT}(r) = A^2 +  [B^2+A(A+1)] \cosech r - B(2A+1) \cosech r \coth r\,.
\ee
whose bound state energy eigenvalues and eigenfunctions are well known. 
Remarkably, if we consider
\be\label{3}
W = W_{GPT} +\frac{2B \sinh r}{2B \cosh r-2A-1}
-\frac{2B \sinh r}{2B \cosh r-2A+1}\,,
\ee
then we find that even though the potential $W^2(r)-W'(r)$ is very different 
from $V_{GPT}$ and given by
\be\label{4}
V(r) = V_{GPT} +\frac{2(2A+1)}{2B\cosh r-2A-1}
-\frac{2[4B^2-(2A+1)^2]}{(2B\cosh r-2A-1)^2}\,,
\ee
the bound state spectrum is still the same and given by
\be\label{5}
E_\nu = A^2 -(A-\nu)^2\,,~~ \nu = 0,1,...,\nu_{max}\,,
\ee
where $A-1 \le \nu_{max} < A$. However, the eigenfunctions are now different
and they are now given by 
\be\label{5a}
\psi_{\nu}(r) = N_\nu \frac{(\cosh r - 1)^{\frac{1}{2}(B-A)} (\cosh r + 1)^{-\frac{1}{2}(B+A)} } {2B\cosh r-2A-1} \hat{P}_{\nu + 1} ^{(\alpha,\beta)}(\cosh r)       \label{B}
\ee
where  $\alpha = B-A-\frac{1}{2}$ , $\beta = -B-A-\frac{1}{2}$. 
Here
\be\label{6}
 N_\nu  = -2^{A+2} B\left(\frac {\nu! (2A-2\nu)(B+A-\nu + \frac {1}{2})
\Gamma(B+A-\nu - \frac{1}{2})}{(B-A+\nu +\frac {1}{2})\Gamma(B-A 
+ \nu-\frac {1}{2})\Gamma(2A-\nu + 1)}\right)^\frac{1}{2}\,,
\ee
 is the 
normalization constant, and  $\hat{P}^{(\alpha , \beta )} _{\nu+1} $ 
is $ (\nu+1)$ th-degree  $X_1 $- Jacobi Polynomial.\\\

 The $X_1$ Jacobi polynomial is related to the usual Jacobi polynomial as \cite{eop2}:
 \be\label{7}
 \hat{P}^{(\alpha , \beta )} _\nu (r) = -\frac{1}{2}(x-b)P^{(\alpha , \beta )} _{\nu-1} (r) + \frac{bP^{(\alpha , \beta )} _{\nu-1} (r) - P^{(\alpha , \beta )} _{\nu-2} (r)}{(\alpha + \beta + 2\nu - 2)}         
 \ee
 where $ b = \frac{\beta + \alpha}{\beta - \alpha} $ \\
 Using (\ref{7}) the $X_1 $ Jacobi Polynomial 
$ \hat{P}^{(\alpha , \beta )} _{\nu+1} (\cosh r)$ can be written as
 \be\label{8}
 \hat{P}^{(\alpha , \beta )} _{\nu+1} (\cosh r) = \frac{1}{2(\alpha + \beta + 2\nu )}\left[\{ (b-\cosh r)(\alpha + \beta + 2\nu )+2b \}P^{(\alpha , \beta )} _{\nu} (\cosh r) - 2P^{(\alpha , \beta )} _{\nu-1} (\cosh r) \right ]   
 \ee 
 Usual Jacobi polynomial $P^{(\alpha , \beta )} _{\nu} (\cosh r)$ further can be written in terms of Hypergeometric function as :\\
 \be\label{10}
P^{(\alpha , \beta )} _{\nu} (\cosh r) = \frac{\Gamma(\nu+\alpha+1)}{ \nu! \Gamma (1+\alpha)} F(\nu+\alpha+\beta+1
, -\nu , 1+\alpha ; \frac{1-\cosh r}{2})\,.
 \ee
 
To get the scattering states for this system two modifications of the bound state wavefunctions have to be made\cite{ks}:
(i) The second solution of the Schr\"odinger equation must be retained - it 
has been discarded for bound state problems since it diverged asymptotically.
(ii) Instead of the parameter $\nu $ labeling the number of nodes, one must use the wavenumber $k$ so that we get the asymptotic 
behavior in terms of $e^{\pm ikr}$ as $r\rightarrow \infty $.
After considering the second solution, Eq. (\ref{10}) becomes
 \ba\label{10b}
P^{(\alpha , \beta )} _{\nu} (\cosh r)& = & \frac{\Gamma (\nu+\alpha+1)}{ \nu ! \Gamma (1+\alpha)} \left[C_1 F(\nu+\alpha+\beta+1
,-\nu , 1+\alpha ; \frac{1-\cosh r}{2})\right. \\ \nonumber 
& + & \left. C_2 (\frac{1-\cosh r}{2})^{-(\nu+\alpha+\beta+1)} \ \ 
F(\nu+\beta + 1, -\nu -\alpha , 1- \alpha ; \frac{1-\cosh r}{2})\right]\,,    
\ea
where $C_1,C_2$ are arbitrary constants.
Considering the boundary condition, i.e as $r\rightarrow 0$ , $(\frac{1-\cosh r}{2}) \rightarrow 0, \psi_{\nu}(r) $ tending to finite,
the allowed solution is
\be\label{11}
P^{(\alpha , \beta )} _{\nu} (\cosh r) = \frac{\Gamma(\nu+\alpha+1)}{ \nu ! \Gamma (1+\alpha)} C_1 F(\nu+\alpha+\beta+1
,-\nu , 1+\alpha ; \frac{1-\cosh r}{2})\,.     
\ee
Now replace $\nu$ by $A+ik$ and use $\alpha+\beta = -2A-1$ We get
\be\label{13}
P^{(\alpha , \beta )} _{(A+ik)} (\cosh r) = C_1\frac{\Gamma(B+ik+1/2)}{ (A+ik) ! \Gamma (B-A+1/2)}F(-A+ik,-A-ik,B-A+1/2; \frac{1-\cosh r}{2})\,.          
\ee
Using  Eq. (\ref{13})  in (\ref{8}) we get $P^{(\alpha , \beta )} _{(\nu+1)} (\cosh r)=P^{(\alpha , \beta )} _{(A+ik+1)} (\cosh r)$.\\
Thus the scattering state wavefunctions becomes
\be\label{15}
\psi_{k}(r) = N_k \frac{(\cosh r - 1)^{\frac{1}{2}(B-A)} (\cosh r + 1)^{-\frac{1}{2}(B+A)} } {2B\cosh r-2A-1} P_{A+ik+1} ^{(\alpha,\beta)}(\cosh r) 
\label{L}
\ee

Now using the properties of hypergeometric function \cite{toi} i.e
\ba\label{16}
F(\alpha,\beta,\gamma;z) &=& (1-z)^{-\alpha} \frac{\Gamma (\gamma) 
\Gamma (\beta - \alpha)}{\Gamma (\beta) \Gamma (\gamma - \alpha)}
F(\alpha,\gamma - \beta, \alpha-\beta+1; \frac{1}{1-z})  \nonumber   \\
&+&(1-z)^-\beta \frac{\Gamma (\gamma) \Gamma (\alpha - \beta)}{\Gamma (\alpha) 
\Gamma (\gamma - \beta)}F(\beta,\gamma - \alpha, \beta-\alpha+1; \frac{1}{1-z})
\ea
and taking the limit $r\rightarrow \infty$ the fourth term of the hypergeometric equation vanishes. Finally we get the asymptotic
form of (\ref{15}), which is given as
\be\label{17}
\lim_{r\to\infty}\psi_{k}(r) = N_k \frac{C_1 2^{-2ik-3A}[(2ik-1)aP+Qc]}
{4B(2ik-1)}\left[\frac{bP(1-2ik)2^{-4ik}}{aP(2ik-1)+Qc} e^{ikr} 
- e^{-ikr}\right]          
\ee
where\\
\[P = \frac{\Gamma(B+ik+1/2)}{ (A+ik) ! \Gamma (B-A+1/2)} ; \ \ Q = \frac{\Gamma(B+ik-1/2)}{(A+ik-1)!\Gamma (B-A+1/2)} ;\] \\  \[a = \frac{\Gamma(B-A+1/2)\Gamma(-2ik)}{\Gamma(-A-ik)\Gamma (B-ik+1/2)}; \ \ b = \frac{\Gamma(B-A+1/2)\Gamma(2ik)}{\Gamma(-A+ik)\Gamma (B+ik+1/2)} ;\]\\
\[c = \frac{\Gamma(B-A+1/2)\Gamma(-2ik+2)}{\Gamma(-A-ik+1)\Gamma (B-ik+3/2)}; \ \ d = \frac{\Gamma(B-A+1/2)\Gamma(2ik-2)}{\Gamma(-A+ik-1)\Gamma (B+ik-1/2)} ;\]\\
The asymptotic behavior for the radial wavefunction (for l=0) is given by 
\cite{cks}
\be\label{18}
\lim_{r\to\infty}\psi_{k}(r) \simeq \frac{1}{2k}[S_{l=0} e^{ikr} 
- e^{-ikr} ]    
\ee
From (\ref{17}) and (\ref{18}) we get
\be\label{19}
S_{l=0} = \frac{bP(1-2ik)2^{-4ik}}{aP(2ik-1)+Qc}                      
\ee

Using P, Q, a, b and c, we get after simple calculation 
(using $\Gamma (n+1) = n\Gamma (n) $)
\ba\label{20}
&&S_{l=0} = S_{l=0}^{GPT} \frac{[B^2-(ik-1/2)^2]}{[B^2-(ik+1/2)^2]} \nonumber \\
&&= \frac{\Gamma(2ik)\Gamma(-A-ik)\Gamma(B-ik+1/2) 2^{-4ik}}
{\Gamma(-A+ik)\Gamma(-2ik) \Gamma(B+ik+1/2)}
\frac{[B^2-(ik-1/2)^2]}{[B^2-(ik+1/2)^2]}\,.
\ea
Thus we notice that the scattering amplitudes for the two potentials (i.e.
GPT and new $X_1$ SIP) are
different even though the bound state spectrum is identical for them. 
It will be interesting to see how the scattering amplitude change as we go 
from the potential here to the one whose eigenfunctions are in terms of
$X_n$ Jacobi polynomials.\\ 
{\bf Acknowledgement}\\
One of us (RKY) acknowledges the financial support from UGC under FIP scheme.

\end{document}